\documentstyle[multicol,aps,epsf]{revtex}
\begin{document}
\title{Domain Number Distribution in the Nonequilibrium Ising Model}
\author{E.~Ben-Naim$\dag$ and P.~L.~Krapivsky$\ddag$}
\address{$\dag$Theoretical Division and Center for Nonlinear Studies, 
Los Alamos National Laboratory, Los Alamos, NM 87545}
\address{$\ddag$Center for Polymer Studies and Department of Physics,
Boston University, Boston, MA 02215}
\maketitle
\begin{abstract}
  We study domain distributions in the one-dimensional Ising
  model subject to zero-temperature Glauber and Kawasaki dynamics. The
  survival probability of a domain, $S(t)\sim t^{-\psi}$, and an
  unreacted domain, $Q_1(t)\sim t^{-\delta}$, are characterized by two
  independent nontrivial exponents. We develop an independent
  interval approximation  that provides close estimates for many
  characteristics of the domain length and number distributions 
  including the scaling exponents. 
\vskip 0.1cm
\noindent
{PACS numbers: 02.50.Ey, 05.40.+j, 82.20.Mj}
\end{abstract}

\begin{multicols}{2}

\section{INTRODUCTION} 

The theory of phase ordering kinetics, or domain coarsening, has
undergone a rapid development in recent years\cite{Bray}.  It has been
established that systems quenched from a homogeneous high-temperature
disordered state to a low-temperature multi-phase state do not order
instantaneously; instead, domains of equilibrium ordered phases form
and grow with time as the system approaches local equilibrium on
larger and larger scales.  Generally, a scale-invariant morphology is
developed at late times, and the network of domains is (statistically)
independent of time when lengths are rescaled by a single
characteristic length scale $L(t)$, the typical domain
size.  This length scale exhibits an algebraic growth with
time, $L(t)\sim t^{\nu}$. However, it was recently realized
that additional scaling laws characterized by nontrivial scaling
exponents exist in such systems. Examples for such decay modes are the
autocorrelation function, $A(L)\sim L^{-\lambda}$\cite{fisher}, and
the fraction of the system still frozen in its initial state,
$P_0(t)\sim t^{-\theta}$\cite{dbg2,kbr}. The latter ``persistence''
probability has since been investigated
theoretically\cite{dbg2,dbg1,dhp,bfk,ms,mbcs,der,lee} and
experimentally\cite{ypms} in spin systems, interacting particles
systems\cite{kbr,cardy,krl,bhm,mc}, Lotka-Volterra models \cite{lpe,fk},
breath figures growth\cite{bou}, foams\cite{ld}, and even simple
diffusion \cite{msbc,w}.

Similar to the domain growth exponent, $\nu$, these additional
exponents are sensitive to the nonequilibrium dynamics followed by the
system, and thus are fundamentally different from their equilibrium
counterparts.  Precisely how many independent hidden exponents does a
coarsening system possesses remains an open question.  In this study, we
establish that at least in one-dimension, additional exponents
describe the survival probability and other more subtle statistical
properties of domains.  We examine systems with short-range
interactions described by a scalar order parameter, namely the 1D T=0
Ising model\cite{Ising} evolving according to nonconserved Glauber
dynamics\cite{glauber} and conserved Kawasaki dynamics\cite{kaw}.

This paper is organized as follows. We first define the domain number
distribution in Sec.~II. In the following section, we review our
results for Glauber spin-flip dynamics where we develop and solve
analytically an Independent Interval Approximation (IIA) that assumes
no correlations between adjacent domains.  The IIA predictions compare
well with Monte Carlo simulations by giving a correct description of the
domain statistics as well as good estimates for the underlying
exponents. In Sec.~IV we show that nontrivial exponents underly the
zero temperature limit of the 1D Ising model with Kawasaki
spin-exchange dynamics as well. The IIA, when carefully modified to
conserved dynamics, turns out to be equally useful in this case. 
Summary and conclusions are given in Sec.~IV. 

\section{Domain Number Distribution}

Although we focus in this study on the Ising model, the statistical
properties of domains we are concerned with are relevant to arbitrary
coarsening processes in one spatial dimension. For example, we ask,
what is the domain survival probability $S(t)$, i.e., the probability
that a domain, initially present at the system at time $t=0$, is still
present at time $t$ (see Fig.~1). We will present theoretical and
numerical evidence supporting an algebraic long time decay of this
survival probability,

\begin{equation}
\label{st}
S(t)\sim t^{-\psi}. 
\end{equation}
Such a behavior is robust, as the exponent $\psi$ is not sensitive to
the initial state of the system (provided long ranged correlations are
absent). Our results will also strongly suggest that the exponent
$\psi$ is nontrivial, i.e., it cannot be extracted from so-far known
exponents associated with the Ising model. 

\begin{figure}
\vspace{-.25in}
\centerline{\epsfxsize=7cm \epsfbox{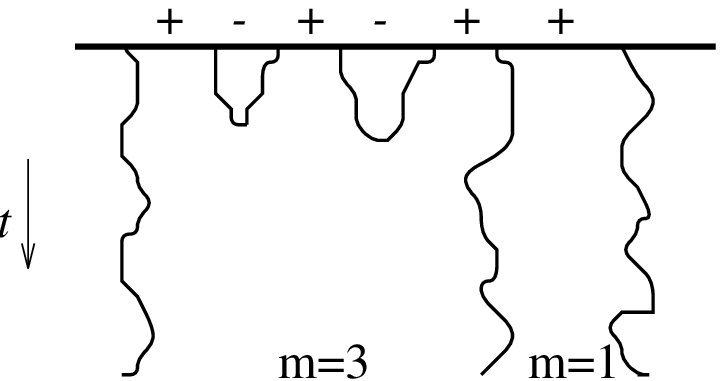}}
\noindent
{\small {\bf Fig.~1}.  
Domain motion in the Ising-Glauber model. Surviving domains 
are marked by $+$, annihilated domains by $-$. The domain number 
at a later time is also indicated.}
\end{figure}

In principle, a surviving domain may undergo coalescence with other
similar phase domains.  Thus, a natural generalization of the domain
survival probability is $Q_m(t)$, the density of domains composed of
$m$ original domains (see Fig.~1).  This quantity satisfies the
initial condition $Q_m(0)=\delta_{m,1}$.  The total domain density,
$N(t)$, is given by $N(t)=\sum_m Q_m(t)$, while the domain
survival probability counts {\em initial} domains that have not shrunk
and hence contains the density $Q_m(t)$ with weight $m$  
\begin{equation}
\label{stdef}
S(t)=\sum_m m Q_m(t).  
\end{equation}
The average number of domains contained within a surviving domain
$\langle m(t)\rangle=S(t)/N(t)$ grows algebraically according to
$\langle m(t)\rangle\sim t^{\nu-\psi}$ with $\nu$ the domain decay
exponent, $N(t)\sim t^{-\nu}$.  If the behavior of $Q_m(t)$ is truly
self-similar, it should follow the scaling form

\begin{equation}
\label{qmscl}
Q_m(t)\simeq t^{\psi-2\nu}{\cal Q}(mt^{\psi-\nu}).
\end{equation}

The scaling function ${\cal Q}(z)$ exhibits the following extremal
behavior
\begin{equation}
\label{qzscl}
{\cal Q}(z)\sim\cases{z^{\sigma}&$z\ll 1$,\cr 
\exp(-\kappa z^{\mu})&$z\gg 1$.}
\end{equation}
The small argument tail describes domains that contain a very small
number of initial domains. In particular, the quantity 

\begin{equation}
\label{qtscl}
Q_1(t)\sim t^{-\delta}
\end{equation}
is of special interest: It gives the density of domains which avoided
merging with their neighboring domains up to time $t$.

The inequalities $Q_1(t)\leq \sum_m Q_m(t)\leq \sum_m m Q_m(t)$ lead to
the bounds $\psi\leq\nu\leq\delta$.  Taking into account that at least
one surviving domain surrounds a persistent spin gives $P(t)\leq S(t)$,
where $P(t)\sim t^{-\theta}$ is the density of persistent spins.  Thus
we arrive at another upper bound $\psi\leq\theta$ for the exponent
$\psi$.  These bounds suggest that the domain decay rate is the slowest
in the problem.  We shall show below that these bounds are strict for
the Ising model and more generally for the $q$-state Potts model.
However, for the Potts model with $q\to 1$ or $q\to\infty$, and for a
few other models\cite{kb} some of these exponents are equal to each other.

A useful relation between the scaling
exponents can be obtained by substituting $m=1$ in Eq.~(\ref{qmscl})

\begin{equation}
\label{sr}
\delta-\nu=(\nu-\psi)(1+\sigma).
\end{equation}
Thus, among the three exponents $\psi$, $\delta$, and $\sigma$, only
two are independent. It is well known that under nonconserved
(conserved) dynamics $\nu=1/2$ ($\nu=1/3$)\cite{Bray}. 

Quite obviously, domains disappear when their size vanishes, and
therefore the domain size and number distributions are intimately
related. Thus, domain survival properties involve the distribution of
domains of size $n$ consisting of $m$ original domains at time $t$,
denoted by $P_{n,m}(t)$. The aforementioned number distribution is
$Q_m(t)=\sum_n P_{n,m}(t)$, and consequently, the domain survival
probability is $S(t)=\sum_{n,m} m P_{n,m}(t)$. 

As will be seen later, studying the joint size-number distribution
requires detailed knowledge of the domain size distribution
$P_n(t)=\sum_m P_{n,m}(t)$. This distribution obeys the normalization
conditions  
\begin{equation}
\label{norm}
1=\sum_n nP_n(t),\qquad N(t)=\sum_n P_n(t).
\end{equation}
Length conservation implies the first relation, while the second
relation gives the total domain density.  Since the average domain
length grows as $n\sim t^{\nu}$, the length distribution follows the
scaling form
\begin{equation}
\label{pnscl}
P_n(t)\simeq t^{-2\nu}{\cal P}(nt^{-\nu}).
\end{equation} 
All of the above scaling behavior emerges from the approximation
detailed below. Furthermore, it is satisfied by the simulation data.
In the next section, we develop an approximation scheme that helps
elucidate many of the qualitative and quantitative features of the
domain size and number distributions.

\section{Nonconserved Glauber dynamics}

We start with the 1D Ising model subject to $T=0$ Glauber dynamics
\cite{glauber}. To examine the role of the number of equilibrium phases
we also consider a generalization of the Ising model, the $q$-state
Potts model. In higher dimensions, the $q$-state Potts model is relevant
to physical situations for $q=2$ (the Ising model) and additionally for
$q=3,4,\infty$\cite{wu}.  For instance, the $q=\infty$ case describes
several cellular structures\cite{st} including polycrystals\cite{poly},
foams\cite{ld}, soap froth\cite{glaz}, and magnetic bubbles\cite{bubbl}.

We consider uncorrelated initial conditions where each of the $q$
phases is present with equal density $1/q$.  The $T=0$ Glauber-Potts
dynamics proceeds by selecting a spin at random and changing its value
to that of one of its randomly selected neighbors.  Thus, domain walls
perform a random walk and upon contact, they annihilate or coalesce,
depending on the state of the corresponding domains
\cite{bbd,racz,amar}. Identifying a domain wall with a particle,
($A$), and absence of a domain wall with a hole ($0$), one finds the
single-species diffusion-reaction process 

\begin{equation}
\label{aa}
A0\buildrel {1\over 2} \over \rightleftharpoons 0A,\quad
AA\buildrel {1\over q-1} \over \longrightarrow  00,\quad
AA\buildrel {q-2\over q-1}\over \longrightarrow A0\ {\rm or }\ 0A. 
\end{equation}
The rates indicate the relative probabilities by which each event 
occurs.

\subsection{Domain Size Distribution}

Ignoring correlations between neighboring domains allows us to develop
an approximate theory for the time-evolution of the domain
distribution. Approximations that are similar in nature proved useful
in studies of related reaction-diffusion processes\cite{msbc,bbd,ab}.

The joint number distribution requires knowledge of the length
distribution and we start by deriving a master equation for $P_n(t)$.
Under the assumption that the lengths of neighboring intervals are
uncorrelated, we write the following rate equation\cite{kb} 
\begin{eqnarray}
\label{pn}
{dP_n\over dt}=&&P_{n-1}+P_{n+1}-2P_n\\
+&&{P_1\over (q-1)N^2}\left[\sum_{i=1}^{n-2} P_i P_{n-1-i}
-N(P_n+P_{n-1})\right].\nonumber
\end{eqnarray}
with $N(t)=\sum_{n}P_n(t)$  the total domain density and the
boundary condition $P_0(t)=0$.  The first three terms reflect that
domain walls perform a random walk with hopping rate set to 1/2
without loss of generality.  The last two terms are due to domain
annihilation: the convolution term accounts for domain merger and the
last term for domain loss. In the $q$-state Potts model collision of
domain walls results in annihilation with probability ${1\over q-1}$
or in coalescence with probability ${q-2\over q-1}$. Only annihilation
events affect the domain distribution and thus the ${1\over q-1}$
prefactor of the annihilation terms.  Using the sum rules of
Eq.~(\ref{norm}), one verifies that the total length is conserved and
the total domain density decays according to the {\it exact} rate
equation

\begin{equation}
\label{N}
{dN\over dt}=-{q\over q-1}P_1.
\end{equation}

The diffusion term in Eq.~(\ref{pn}) implies $\langle n(t)\rangle\sim
t^{1/2}$, and since $\langle n\rangle\sim N^{-1}$ the correct decay
exponent $\nu=1/2$ \cite{glauber} is recovered.  In the following, we
will need to determine the asymptotic prefactor $A$, $N(t) \simeq
At^{-1/2}$, $A=\int dx\,{\cal P}(x)$, with the scaling function ${\cal
  P}(x)$ defined according to Eq.~(\ref{pnscl}).  The density rate
equation (\ref{N}) implies $P_1\simeq {\cal P}'(0)t^{-3/2}$ with
${\cal P}'(0)={q-1\over 2q}A$.

A quantitative analysis of Eq.~(\ref{pn}) may be carried by treating
the variable $n$ as continuous. The quantity ${\cal P}(x)$ satisfies

\begin{equation}
\label{px}
{\cal P}''+{1\over 2}(x{\cal P})'+{q-2\over 2q}{\cal P}
+{1\over 2qA}\,{\cal P}*{\cal P}=0,
\end{equation}
where ${\cal P}'\equiv d{\cal P}/dx$ and ${\cal P}*{\cal
P}\equiv\int_0^x dy\,{\cal P}(y){\cal P}(x-y)$. The 
normalized Laplace transform of the scaling function ${\cal P}(x)$, 
$p(s)=A^{-1}\int_0^{\infty} dx\,e^{-sx}\,{\cal P}(x)$, obeys 

\begin{equation}
\label{riccati}
{dp\over ds}={p^2\over qs}+\left(2s+{q-2\over qs}\right)p
-{q-1\over qs},
\end{equation}
subject to the boundary condition $p(0)=1$.  The transformation
$p(s)=1-qs^2-qs{d\over ds}\ln y(s)$ reduces the Riccati equation
(\ref{riccati}) into the parabolic cylinder equation,

\begin{equation}
\label{par}
{d^2 y\over ds^2}+\left(1+{2\over q}-s^2\right)y=0. 
\end{equation}
The solution to (\ref{par}) reads $y(s)=C_-D_{1/q}(-s\sqrt{2})+C_
+D_{1/q}(s\sqrt{2})$, with $D_{1/q}(x)$ the parabolic cylinder function
of order $1/q$ \cite{bo}.  The large $s$ behavior of $p(s)$, $p(s)\simeq
{q-1\over 2q}s^{-2}$, implies $C_-=0$, and we get

\begin{equation}
\label{ps}
p(s)=1-qs^2-qs{d\over ds}\ln D_{1/q}(s\sqrt{2}).
\end{equation}
The normalization condition $\sum_n nP_n(t)=1$ can be
reduced to $Ap'(0)=-1$.  This allows us to determine the constant

\begin{equation}
\label{A}
A={\Gamma\big[1-{1\over 2q}]
\over\Gamma\big[{1\over 2}-{1\over 2q}\big]},
\end{equation}
where $\Gamma$ denotes the gamma function.  In deriving (\ref{A}) 
we have used the properties \cite{bo}

\begin{equation}
\label{large}
D_c(x)\sim x^c\exp(-x^2/4)[1+{\cal O}(x^{-2})],
\end{equation}
and
\begin{equation}
D_c(0)={\pi^{1/2}2^{c/2}\over \Gamma(1/2-c/2)},\quad
D_c'(0)=-{\pi^{1/2}2^{(c+1)/2}\over \Gamma(-c/2)}.
\end{equation}

The value of the constant $A$ predicted by the IIA may be compared to
the exact one, $A_{\rm exact} =(1-q^{-1})/\sqrt{\pi}$ \cite{amar} In
the extreme cases of $q=1$ and $q=\infty$ the prefactor $A$ is exact.
The mismatch is worst (roughly 20\%) for the Ising ($q=2$) case where
$A=\Gamma(3/4)/\Gamma(1/4)\cong 0.337989$ while $A_{\rm exact} 
=(4\pi)^{-1/2}\cong 0.28209$\cite{glauber}.

The IIA predicts the correct qualitative behavior of length
distribution in the limits of small and large intervals 

\begin{equation}
\label{pxscl}
{\cal P}(x)\sim\cases{{A(q-1)\over 2q}x&$x\ll 1$,\cr 
qA \exp(-\lambda x)&$x\gg 1$.}
\end{equation}
The linear small size behavior is seen from the large $s$ behavior
$p(s)\simeq {(q-1)\over 2q}s^{-2}$.  On the other hand, the
exponential tail follows from the behavior of the Laplace transform
$p(s)\simeq q\lambda/(s+\lambda)$ near its pole at negative
$s=-\lambda$, given by the first zero of
$D_{1/q}(-\lambda\sqrt{2})=0$.  For the Ising case one has
$\lambda=0.5409$. This value should be compared with the exact value
$\lambda=\zeta(3/2)/4\sqrt{\pi} =0.368468$ obtained by Derrida and
Zeitak\cite{dz} and the approximate value $\lambda=0.35783$ obtained
by Alemany and ben-Avraham\cite{ab}.

\subsection{Domain Size-Number Distribution}

We are now in a position to tackle the joint size-number distribution,
$P_{n,m}(t)$, which captures both the spatial and ``historical''
characteristics of the coarsening domain mosaic.  The corresponding
rate equation is a generalization of Eq.~(\ref{pn})

\begin{eqnarray} 
\label{pnm}
&&{dP_{n,m}\over dt}=P_{n-1,m}+P_{n+1,m}-2P_{n,m}\\
&&+{P_1\over (q-1)N^{2}}\left[\sum_{i,j}P_{i,j}P_{n-1-i,m-j}
-N(P_{n,m}+P_{n-1,m})\right] \nonumber
\end{eqnarray}
with the initial condition $P_{n,m}(0)=\delta_{n,1}\delta_{m,1}$ and the
boundary condition $P_{0,m}(t)=0$.  The variable $m$ is almost mute as it
appears in a nontrivial way only in the convolution term.  One should
verify that this master equation is self-consistent.  First, by summing
over $m$, we recover Eq.~(\ref{pn}).  Second, it implies that the domain
survival probability satisfies the exact linear equation
${dS/dt}=-\sum_m m P_{1,m}$.  

We have not succeeded in solving for the joint distribution.
Nevertheless, it is possible to obtain analytically many interesting
properties of Eqs.~(\ref{pnm}), including the scaling exponents.  Given
Eqs.~(\ref{pnm}) are recursive in $m$, one can try to solve for
$P_{n,1}(t)$, then for $P_{n,2}(t)$, etc.  A solution for the former
quantity already allows to determine the scaling exponent $\delta$.
Thus let us consider the distribution of domains which have not merged
with other domains up to $t$, $R_n(t)\equiv P_{n,1}(t)$. For such
domains, the convolution term vanishes and they evolve according to the
linear rate equation
\begin{equation}
\label{rn}
{dR_n\over dt}=R_{n-1}+R_{n+1}-2R_n-{P_1\over (q-1)N}(R_n+R_{n-1})
\end{equation}
with the initial condition $R_n(0)=\delta_{n,1}$ and the boundary
condition $R_0(t)=0$.  In the continuum limit we again replace
$R_{n-1}+R_{n+1}-2R_n$ by $\partial^2 R/\partial n^2$ and
$R_n+R_{n-1}$ by $2R_n$ to find a diffusion-convection equation for
$R_n(t)$.  The transformation $R_n\to \tilde R_n N^{2/q}$ reduces this
equation to the diffusion equation for $\tilde R_n$, which is solved to
yield $R_n(t)\simeq N^{2/q}t^{-1}{\cal R}(nt^{-1/2})$, with ${\cal
R}(x)=x\exp(-x^2/4)/\sqrt{\pi}$.  The total density of unreacted domains
is $Q_1(t)=\sum_n R_n\sim t^{-{1\over 2}-{1\over q}}$, which gives the
decay exponent
\begin{equation}
\label{delta}
\delta={1\over 2}+{1\over q}.  
\end{equation}

Obtaining the second independent exponent $\psi$ is more involved.  The
natural approach, i.e., a direct investigation of the domain number
distribution $Q_m$, appears to be useless, as it requires knowledge of
$P_{1,m}$ and hence the entire $P_{n,m}$. The domain survival
probability can be alternatively obtained by considering $U_n(t)=\sum_m
mP_{n,m}(t)$. This quantity obeys

\begin{eqnarray}
\label{un}
{dU_n\over dt}&=&U_{n-1}+U_{n+1}-2U_n\\
&+&{P_1\over (q-1)N^2}
\left[2\sum_{i=1}^{n-2} U_i P_{n-1-i}-N(U_n+U_{n-1})\right],\nonumber
\end{eqnarray}
obtained by summing Eqs.~(\ref{pnm}).  We write $U_n(t)$ in a scaling
form $U_n(t)\simeq t^{-\psi-1/2}{\cal U}(nt^{-1/2})$.  Asymptotically,
the domain survival probability reads $S(t)\simeq Bt^{-\psi}$ with
$B=\int dx\, {\cal U}(x)$.  The scaling distribution satisfies

\begin{equation}
\label{ux}
{\cal U}''+{1\over 2}(x{\cal U})'+ \left(\psi-{1\over
q}\right){\cal U} +{1\over qA}{\cal U}*{\cal P}=0.  
\end{equation}
The  normalized Laplace transform of the scaling function ${\cal U}(x)$,
$u(s)=B^{-1}\int_0^\infty dx\,e^{-sx}{\cal U}(x)$, obeys

\begin{equation}
\label{dus}
{du\over ds}=2\left({p(s)+q\psi-1\over qs}+s\right)u
-{2\psi\over s}, 
\end{equation}
and $u(0)=1$.  In deriving (\ref{dus}) we used the relation ${\cal
  U}'(0)=B\psi$, found by integration of Eq.~(\ref{ux}), combined with
$A=\int dx {\cal P}(x)$. Substituting the explicit expression
(\ref{ps}) for $p(s)$ into Eq.~(\ref{dus}), and solving for $u(s)$
yields
\begin{equation} 
\label{us} u(s)=2\psi
s^{2\psi}D^{-2}_{1/q}(s\sqrt{2}) \int_s^{\infty}\!\!dr\,
r^{-2\psi-1}D_{1/q}^{2}(r\sqrt{2}).  
\end{equation} 
This solution is consistent with the anticipated $s\to\infty$ behavior,
$u(s)\simeq \psi s^{-2}$. Furthermore, evaluating Eq.~(\ref{us}) near
the origin gives $u(s)=1+F(\psi)s^{2\psi}+Cs+\cdots$.  Therefore, for
$u'(s)$ to be finite near $s=0$, we must have $F(\psi)=0$. Evaluating
$F(\psi)$ gives

\begin{equation} 
\label{psi} 
0=\int_0^\infty dr\,r^{-2\psi}D_{1/q}(r)D_{1/q}'(r),
\end{equation} 
an eigenvalue problem that can be solved numerically to obtain the
exponent $\psi$ (see Table 1).  In the most interesting case of integer
$q$ the domain decay exponent $\psi$ appears to be irrational, in
contrast with $\delta$.

It is useful to consider the limiting cases that turn out to be
solvable.  The $q=\infty$ limit is especially simple\cite{bbd} as only
domain walls coalesce but cannot annihilate and thereore Eq.~(\ref{pnm})
is linear and thus exact.  Furthermore, the domain size number
distribution factorizes, $P_{n,m}=P_n(t)\delta_{m,1}$, since similar
phase domains never coalesce and therefore the domain number is trivial,
$m=1$.  Thus $N(t)=S(t)=Q_1(t)\simeq (\pi t)^{-1/2}$ and
$\nu=\psi=\delta=1/2$.  Additionally, the scaling function is ${\cal
P}(x)=x\exp(-x^2/4)/\sqrt{\pi}$.

Before going to the opposite limit $q\to 1$, we first note that the
Potts model with arbitrary $q\geq 1$ can be mapped onto the Ising model
with magnetization $\mu=2/q-1$.  Thus the $q\to 1$ limit corresponds to
the vanishing volume fraction of minority domains.  Therefore minority
domains cannot ``meet'', so majority domains change appreciably only due
to coalescence.  Thus they never disappear, i.e., $S(t)=1$ and $\psi=0$.
A majority domain remains unreacted if both its minority neighbors
survive, implying $Q_1(t)=N^2(t)$ and $\delta=2\nu=1$.  In the above
rate equation description, the diffusion term becomes negligible and the
IIA is exact.  The scaling functions ${\cal P}(x)$ and ${\cal Q}(z)$ are
identical exponential functions.  However, unlike to the $q=\infty$
case, the joint distribution is not a product of the single function
variables\cite{kb}.  The joint size-number distribution still obeys the
scaling law, $P_{n,m}(t)\sim t^{-5/4}\Phi(x,y)$, with
$\Phi(x,y)=x^{-1/2}\exp(-x-y^2/2x)$.  The scaling variables $x$ and $y$
are quite unusual, though\cite{aggr,kb}.  Indeed, instead of the naive scaling
variables $mt^{-1/2}$ and $nt^{-1/2}$, one has $x=(m+n)(\pi t)^{-1/2}$
and $y=(m-n)(\pi t)^{-1/4}$\cite{kb}.  The former scaling variable $x$
is just the sum of the naive scaling variables, while the latter
``diffusive'' scale $y$ is hidden.  This suggests that generally for the
$q$-state Potts model with $q<\infty$ the joint distribution is not
necessarily the product of single variable functions and the scaling, if
holds, may be rather different from the naive form with scaling
variables $n t^{-\nu}$ and $m t^{\psi-\nu}$.

\subsection{Simulation Results}

To test the IIA predictions, we performed numerical simulations on a
spin chain of size $L=10^7$.  Random initial conditions and periodic
boundary conditions were used.  The simulation data represents an
average over 10 different realizations. For the Ising case, we found
the exponent values $\psi=0.126(1)$ and $\delta=1.27(2)$ (see Fig. 2).
These values should be compared with the IIA predictions of
$\psi=0.136612$ and $\delta=1$. The IIA neglects correlations that do
build up between neighboring domains and thus is not exact.
Furthermore, the effects of the correlations is nontrivial, as one
exponent is smaller than predicted while the other is larger.  As was
the case for the persistence exponent, $\theta$, the domain exponents
strongly depend on $q$.  Numerical values of the exponents $\psi$ and
$\delta$ are summarized in Table 1 for representative values of $q$.
As $q$ increases, the approximation improves and eventually becomes
exact for the extreme case $q=\infty$.  Thus, $\psi$ is overestimated
by up to 10\% and $\delta$ is underestimated by up to 25\%.

We performed several checks to verify that the asymptotic behaviors of
Eqs.~(\ref{st}) and (\ref{qtscl}) are robust.  For example, they are
independent of the initial domain wall concentration (provided that the
correlations in the initial condition are short range). We conclude that
$\psi$ and $\delta$ are nontrivial exponent, i.e., they cannot be
extracted from the known exponents associated with the Ising-Glauber
model.  Similar to the persistence exponent, $\theta(q)$, the exponents
appear to be irrational except for the limiting cases $q=\infty$
($\psi=\delta={1\over 2}$, $\sigma=0$) and, maybe, for $q=2$
($\psi={1\over 8}$, $\delta={5\over 4}$, $\sigma=1$).

The numerical simulations also confirm that the distribution function
$Q_m(t)$ scales according to Eq.~(\ref {qmscl}). The scaling function
${\cal Q}(z)$, defined in Eq.~(\ref{qzscl}), decays exponentially for
large argument ($\mu=1$) and is algebraic for small argument. The
scaling relations combined with the simulation values give
$\sigma=1.05(5)$.  This is consistent with the linear behavior seen in
our simulations for $z\ll 1$.  Comparing with Eq.~(\ref{pxscl}), we
conclude that similar scaling functions underlie the domain number and
size distributions in the $q=2$ case.

\begin{figure}
\vspace{-.40in}
\centerline{\epsfxsize=9cm \epsfbox{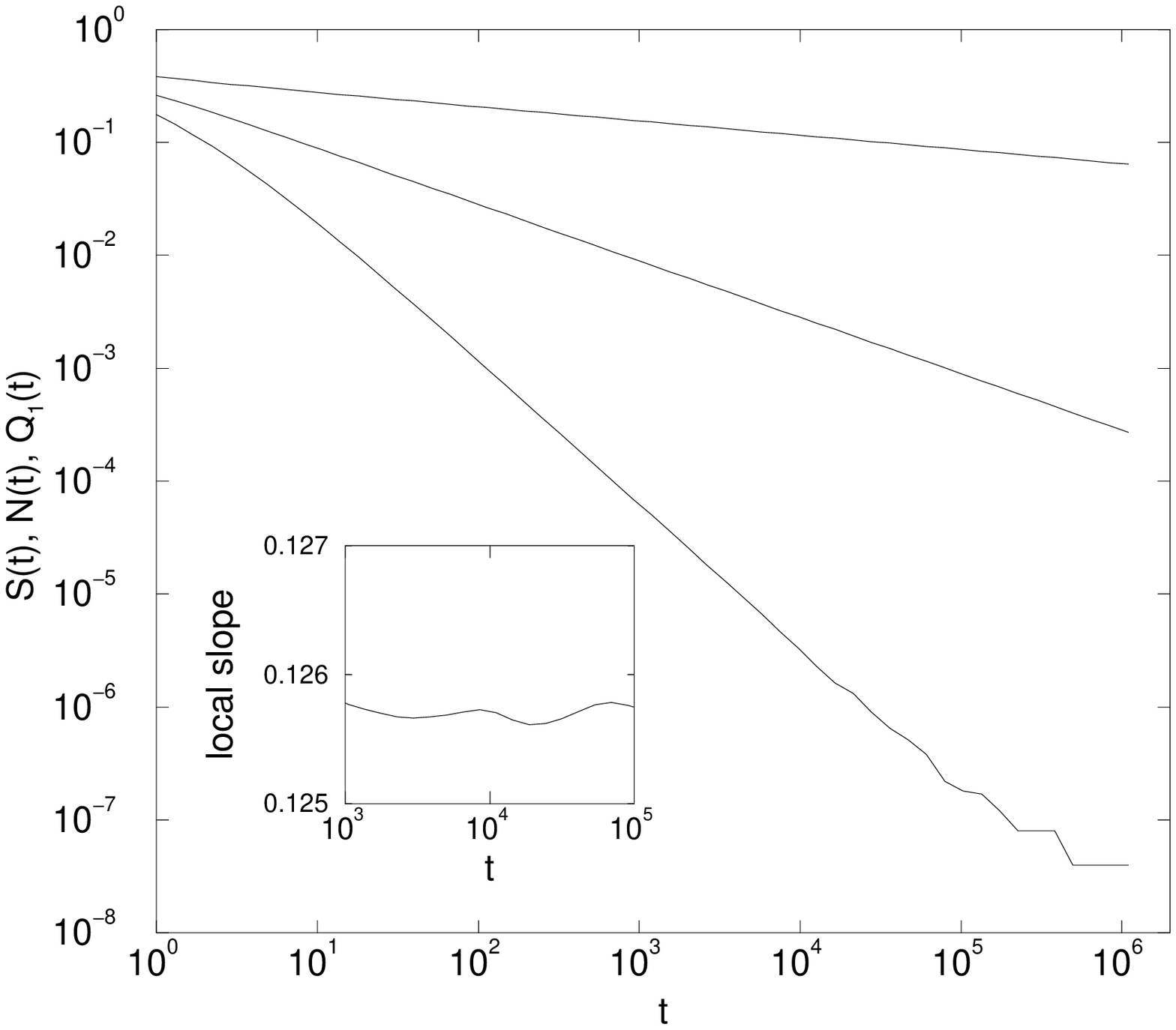}}
\vspace{-.40in} 
\noindent
{\small {\bf Fig.~2}. Monte Carlo data for the Ising-Glauber model.
  The domain survival probability $S(t)$,
  the domain density $N(t)$, and the density of unreacted domains 
  $Q_1(t)$ are shown (top to bottom). The inset plots the local
  slope $-d\ln S(t)/d\ln t$.
  Typically, it is stable over a large temporal range, and thus can be
  used to find the scaling exponents and to estimate the error,
  typically of the order $0.001$.}
\end{figure}

On the other hand, direct numerical integration of Eq.~(\ref{pnm})
reveals a number distribution, $Q_m(t)$, that scales according to
Eq.~(\ref{qmscl}), and has an exponential tail in agreement with the
simulation results. Moreover, the emerging $S(t)$ falls within $5\%$
of the actual survival probability over a significant temporal range,
$t<10^3$.  In summary, in addition to predicting the correct scaling
behavior, Eq.~(\ref{pnm}) provides a good approximation for many
quantitative features of the domain distribution, and in particular,
good estimates for the decay exponents.

\vspace{.1in} 
\centerline{
\begin{tabular}{|l|c|c|c|c|c|}
\hline
&\multicolumn{3}{c|}{MC}&\multicolumn{2}{c|}{Eq.~(\ref{pnm})}\\
\hline
$q$&$\psi$&$\delta$&$\sigma$&$\psi$&
$\delta$\\
\hline
2        &\,0.126\,&\,1.27\,&\,1.05\,&\,0.136612\,&\,1\,\\
3        &\,0.213\,&\,0.98\,&\,0.67\,&\,0.231139\,&\,5/6\,\\
4        &\,0.267\,&\,0.85\,&\,0.50\,&\,0.287602\,&\,3/4\,\\
8        &\,0.367\,&\,0.665\,&\,0.24\,&\,0.385019\,&\,5/8\,\\
50       &\,0.476\,&\,0.525\,&\,0.03\,&0.480274 &13/25\\
$\infty$   &\,  1/2\,&\,1/2 \,&\,0\,&\,  1/2\,&1/2\\
\hline
\end{tabular}}
\vspace{.1in} 
\noindent{\small {\bf Table 1}: Domain exponents for the $q$-state
  Potts model in one dimension. Local slopes analysis was applied to
  the simulation data.  The theoretical
  $\psi$ is from Eq.~(\ref{psi}) and $\delta={1\over 2}+{1\over q}$.}

\section{Conserved Kawasaki dynamics}

We turn now to applying the above methods to the conserved counterpart,
the spin-exchange Kawasaki dynamics\cite{kaw}, which describe spinodal
decomposition in binary alloys and phase separation in binary
liquids.  Although some qualitative features are
known\cite{tob,cor,huse}, theoretical understanding of the
Ising-Kawasaki model is still far from complete even in one dimension.
In the following we limit ourselves to the two-phase Ising case.

\subsection{Reduction to Domain Diffusion}

We start by formulating the appropriate zero-temperature limit of the
Ising-Kawasaki model.  Consider a two-phase system, e.g., the Ising
model (spins up and down) or a binary alloy (atoms of type $A$ and $B$).
At zero temperature energy raising transitions are forbidden and only
two moves are allowed: the energy-decreasing ``coarsening'' transitions
$ABAB\to AABB$ and the energy-conserving ``diffusion'' transitions
$ABAA\to AABA$.  This dynamics ultimately drives the system to a frozen
configuration consisting of strings of alternating domains each of
length $\geq 2$ which could not evolve further\cite{zero}.  This
``jamming'' behavior arises from the nonergodic nature of the
zero-temperature Kawasaki dynamics and it is a robust one: It is
independent of the relative transition rates \cite{cor,anal}, as well as
the spatial dimension \cite{zero,inf,bin}.

Thus, to recover sensible {\em coarsening}, one must consider the
zero-temperature limit. Let us assume that temperature is positive;
when it is sufficiently low, $T\ll J$ where $J$ is the exchange
coupling, the correlation length $\xi\sim e^{J/T}$ is very large and
therefore the system exhibits coarsening as long as the mean domain
size is small compared with the correlation length.  Below, we focus
on this intermediate-time regime where the description is drastically
simplified\cite{tob,cor}.  We assume that the initial stage has been
already completed so that single-spin domains disappeared.  Coarsening
will occur only when a spin splits off a domain wall and penetrates a
neighboring domain (say of size $L$).  The splitting process occurs
with a very small rate $\exp[-4J/T]$.  Then, this spin diffuses inside
the domain until it is eventually adsorbed by its boundaries.  The
corresponding probabilities are well-known from elementary probability
theory\cite{feller}.  The spin will be absorbed by the boundary from
which it was issued with probability $1-1/L$.  This spin may also be
absorbed by the opposite boundary resulting in a one lattice site hop
of the entire domain.  Thus, the hopping rate is $L^{-1}\exp[-4J/T]$.
Rescaling time, $t\to t\exp[-4J/T]$, the spin diffusion will proceed
with a huge rate $\exp[4J/T]$ and it therefore may be treated as
instantaneous while domain hops proceed with a finite rate reciprocal
to the domain size.

Thus, the appropriate zero-temperature limit of the Ising-Kawasaki
model is realized by taking the limits of infinite ``physical'' time
$t_{\rm phys}\to\infty$, while keeping the modified time $t=t_{\rm
  phys}\exp[-4J/T]$ finite.  Hence, entire domains perform a random
walk with rate inversely proportional to their lengths (we ignore an
anomaly concerning domains of length 2 as it is irrelevant
asymptotically).  Heuristically, it may be argued that as diffusion is
the primary coarsening mechanism, the following scaling for the
average domain size, $L\sim \sqrt{Dt}$, holds.  However, since the diffusion
coefficient and the domain size are reciprocal, $D\sim L^{-1}$, we
obtain $L\sim t^{1/3}$ in agreement with the well-known behavior of
systems with conserved scalar order parameter\cite{Bray}.

\subsection{Domain Size Distribution}

For simplicity, we consider the case where the two equilibrium phases
are equivalent, as is the case for random initial conditions.
Modifying Eq.~(\ref{pn}) to account for domain diffusion, and assuming
neighboring domains are uncorrelated, the domain size distribution
evolves according to
\begin{eqnarray}
\label{pn1}
{dP_n\over dt}&=&L^{-1}(P_{n-1}-2P_n+P_{n+1})\\
&+&{P_1\over N^2}\left[\sum_{i+j=n}i^{-1}P_i P_j
-N(n^{-1}+L^{-1})P_n\right],\nonumber
\end{eqnarray}
with the domain density $N$ and the inverse average domain size $L^{-1}$
defined via
\begin{equation}
\label{nln}
N=\sum_{n=1}^\infty P_n, \quad
L^{-1}=\langle
n^{-1}\rangle ={\sum_{n} n^{-1}P_n\over
\sum_{n} P_n}.  
\end{equation}
The diffusion term in Eq.~(\ref{pn1}) accounts for change in $n$ due to
hoping of a neighboring interval.  The convolution term accounts for
gain due to domain merger, and the last two terms represent loss due to
domain collision or domain merger. This rate equation conserves the
total length, $\sum_n nP_n=1$, and summation over $n$ shows that the
total domain density evolves according to $\dot N=-2L^{-1}P_1$.

The length distribution scales according to Eq.~(\ref{pnscl}),
$P_n(t)\simeq t^{-2\nu}{\cal P}(nt^{-\nu})$, with the correct scaling
exponent $\nu=1/3$.  By inserting that scaling form into Eq.~(\ref{pn1})
we arrive at an integro-differential equation for ${\cal P}(x)$ which is
very cumbersome as it involves yet unknown moments of the distribution.
We thus resort to numerical integration of Eq.~(\ref{pn1}).  The results
compare well with Monte-Carlo simulations of the Ising-Kawasaki model
(see Fig.3). For example the estimate for the asymptotic prefactor $A$
(defined via $N(t)\simeq At^{-1/3}$), falls within roughly 5\% of the
actual value: $A_{\rm MC}=0.441\pm0.001$ while $A_{\rm
IIA}=0.415\pm0.005$.

\begin{figure}
\vspace{-.30in}
\centerline{\epsfxsize=9cm \epsfbox{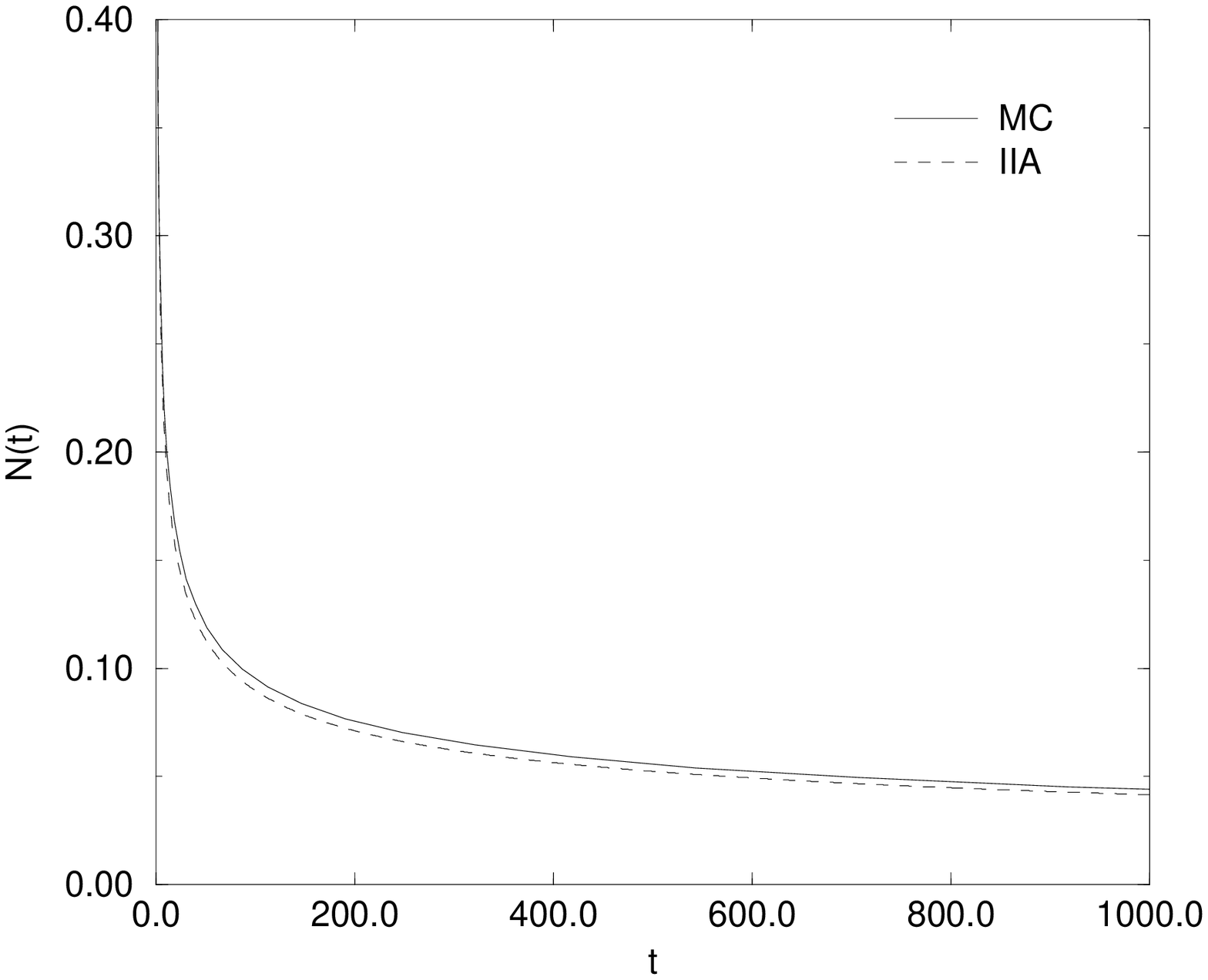}}
\vspace{-.30in} 
\noindent
{\small {\bf Fig.~3}. 
Domain density in the Ising-Kawasaki model. The Monte Carlo simulation
data (MC) represents an average over 10 systems of size $10^5$. The 
IIA was obtained by  integrating  Eq.~(\ref{pn1}) numerically.}
\end{figure}

The length distribution emerging from the IIA has the same limiting
behavior as in the nonconserved case, i.e., it is linear at small $n$
and exponential at large $n$. While the former agrees with our
simulation results, there is a disagreement for the latter.  Our data
is consistent with a Gaussian tail, i.e., ${\cal P}(x)\sim\exp(-x^2)$,
for $x\gg 1$.

\subsection{Domain Size-Number Distribution}

Given the results in the conserved case, it is natural to study the
domain exponents and to examine the usefulness of the IIA approach in
the conserved dynamics case.  In analogy with Eq.~(\ref{pnm}), the
master equation for the domain size-number distribution is written
\begin{eqnarray}
\label{pnm1}
&&{dP_{n,m}\over dt}=L^{-1}(P_{n-1,m}-2P_{n,m}+P_{n+1,m})\\
&+&{P_1\over N^2}\left[\sum_{i+j=n}\sum_{k+l=m}i^{-1}P_{i,k}P_{j,l}
-N(n^{-1}+L^{-1})P_{n,m}\right].\nonumber
\end{eqnarray}
Summing the above equations over $m$, we indeed recover
Eq.~(\ref{pn1}) for the length distribution.  

To determine the exponents, it is again simpler to consider the
distributions $R_n(t)$ and $U_n(t)$ instead of the joint distribution
$P_{n,m}(t)$.  The density of single parent domains, $R_n(t)\equiv
P_{n,1}(t)$, evolves according to the linear rate equation similar to
Eq.~(\ref{rn})
\begin{eqnarray}
\label{rn1}
{dR_n\over dt}&=&L^{-1}(R_{n+1}+R_{n-1}-2R_n)\\
&&-{P_1\over N}(n^{-1}+L^{-1})R_n.\nonumber
\end{eqnarray}
We expect that the distribution $R_n(t)$ scales according to $R_n(t)\sim
t^{-\delta-1/3}{\cal R}(nt^{-1/3})$.  Integrating this equation we get
$Q_1(t)=\sum R_n(t)\sim t^{-\delta}$ with $\delta\cong 0.645$.  On the
other hand, Monte-Carlo simulations of the domain diffusion process give
$\delta\cong 0.705$ (see Fig.~4).

The domain survival probability can be found by using the auxiliary
function $U_n(t)=\sum_m mP_{n,m}(t)$ which satisfies the analog of
Eq.~(\ref{un}) 
\begin{eqnarray}
\label{un1}
{dU_n\over dt}&=&L^{-1}(U_{n+1}+U_{n-1}-2U_n)\\
&&+{P_1\over N^2}
\left[2\sum_{i+j=n}i^{-1}P_iU_j-N(n^{-1}+L^{-1})U_n\right].\nonumber
\end{eqnarray}
This distribution $U_n(t)$ should scale according to $U_n(t)\sim
t^{-\psi-1/3}{\cal U}(nt^{-1/3})$.  The domain survival probability 
$S(t)=\sum_n U_n(t)$ then decays according to $S(t)\sim t^{-\psi}$.

\begin{figure}
\vspace{-.30in}
\centerline{\epsfxsize=9cm \epsfbox{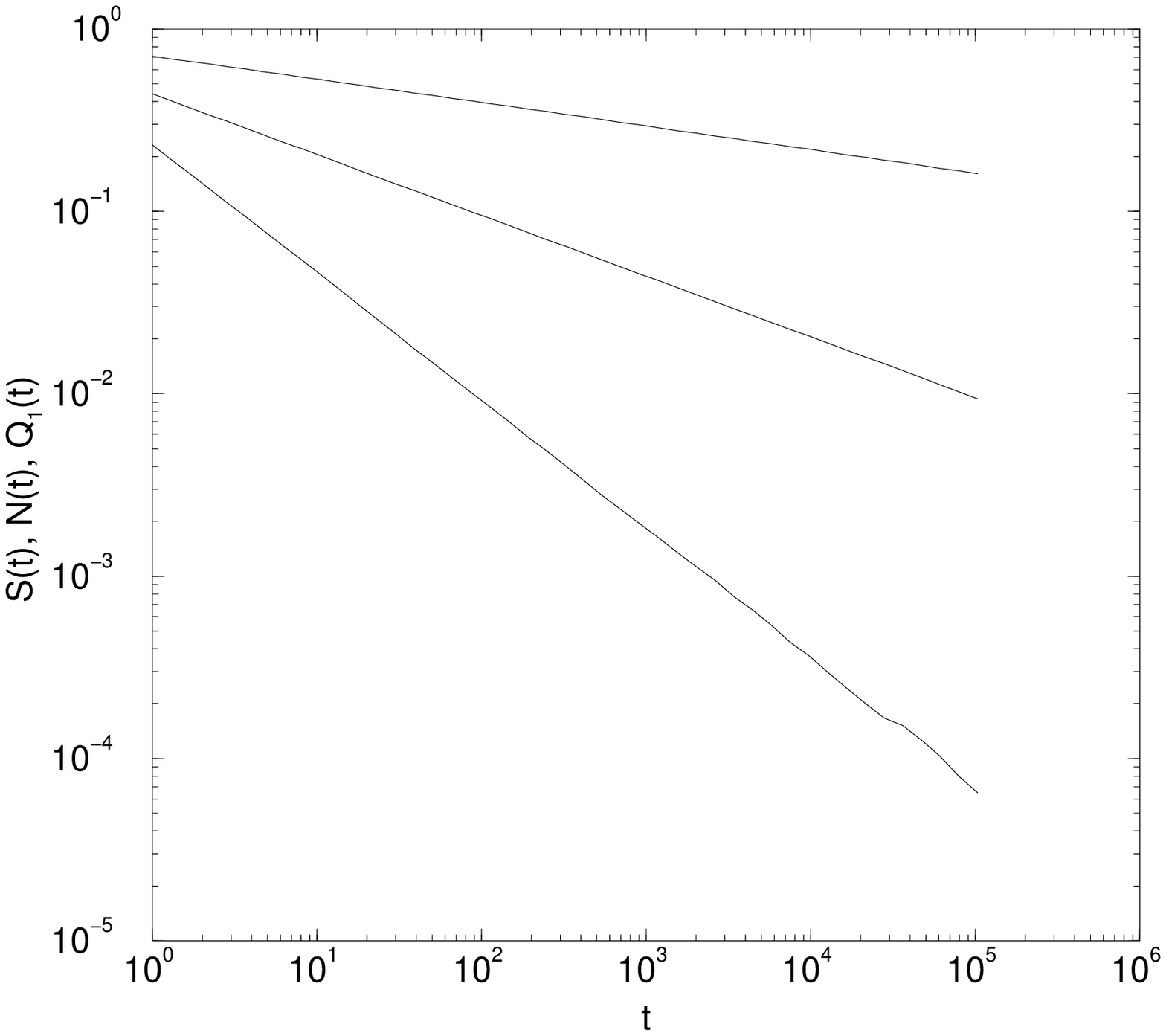}}
\vspace{-.30in} 
\noindent
{\small {\bf Fig.~4}.  Monte Carlo data for the Ising-Kawasaki model
  from the same simulation as in Fig.~(3).  The domain survival
  probability $S(t)$, the domain density $N(t)$, and the density of
  unreacted domains $Q_1(t)$ are shown (top to bottom).}
\end{figure}

Again, the agreement with the simulation is remarkable.  Numerical
integration data give an estimate of $\psi\cong 0.147$ while
Monte-Carlo simulations (see Fig.~4)) give $\psi\cong 0.130$.  We also
verified that the scaling relations of Eqs.~(\ref{qmscl}) and
(\ref{qzscl}) are satisfied by the IIA as well as the simulation
data. We conclude that in the Kawasaki case as well, nontrivial exponents
characterize domain statistics.  Furthermore, the approximate approach
reproduces most qualitative features of the domain size and number
distribution, and provides good estimates for the scaling exponents.

We now describe how to obtain $\psi$ in the limit where one of the two
phases occupies a vanishing volume fraction.  Denote by $L_A(t)$ and
$L_B(t)$ the average sizes of minority and majority domains,
respectively.  They both grow as $t^{1/3}$ but remain greatly
different throughout the evolution, $L_A\ll L_B$.  The domains
diffusion rates, $D_A\sim L_A^{-1}$ and $D_B\sim L_B^{-1}$, thus
greatly differ as well: $D_A\gg D_B$.  In principle, two neighboring
minority domains can overtake a separation distance of order $L_B$ and
coalesce; this requires the coalescence time $t_c\sim L_B^2/D_A\sim
L_AL_B^2$.  On the other case, a minority domain can shrink due to
diffusion of neighboring majority domains; this requires the shrinking
time $t_s\sim L_A^2/D_B\sim L_A^2L_B$.  We see that $t_s\ll t_c$, so
we should just take into account the shrinking of minority domains.

Thus majority domains do not disappear, implying $S(t)=1$ and $\psi=0$.
For the minority phase we anticipate $\psi=\nu=1/3$, while for the
symmetric case of equal concentrations $\psi\cong 0.130$. This indicates
that similar to the nonconserved dynamics case, the domain exponents
vary continuously as the volume fraction is varied.

\section{Conclusions}

In summary, we investigated the one-dimensional Ising model subject to
zero temperature Glauber and Kawasaki dynamics.  We introduced the
domain size-number distribution and showed that it obeys scaling and is
characterized by two independent nontrivial decay exponents. Similar to
the persistence exponent, these exponents are sensitive to the type of
the dynamics and the volume fraction of the (globally or locally)
conserved equilibrium phase. We also introduced an approximation which
is based on terminating the hierarchy of rate equations describing the
domain density. This approximation is very useful in predicting the
qualitative nature of the domain distribution as well as estimating
important parameters including the scaling exponents. In the proper T=0
limit of the Ising model with Kawasaki dynamics, this approximation is
especially useful as very little is known analytically about the domain
distribution.

It will be interesting to generalize the domain survival concept to
higher dimensions. At least for the $q\to\infty$ limit of the Potts
model, domains are well defined, and such a generalization is possible.
The nonconserved dynamics can indeed be studied in this limit, and the
domain exponents $\psi=\theta=\delta=d/2$ (for $d\geq 2$) have been
reported\cite{kb}, consistent with simulations\cite{ld} and with
experiments on $d=2$ soap froths\cite{tam}.

Recently, it was pointed out that coarsening mosaics may be
characterized by more than one algebraically growing length scale, and
that morphologies consisting of domain and super-domains may exist
\cite{lpe}. This, together with the above results suggest that our
current understanding of such systems is only partial.  Domain
statistics indicates that several nontrivial decay laws underlie the
evolution of elementary processes such as the nonequilibrium Ising
model.  These nontrivial exponents do not emerge naturally from
studies of traditional quantities such as spatiotemporal correlations.
It remains a challenge to find and obtain these underlying ``hidden''
exponents from a more systematic method.  It is also intriguing 
whether an entire hierarchy or a finite number of independent decay
modes are present in these systems. 

\bigskip\noindent
This paper is dedicated to Leo Kadanoff on the occasion of
his 60th birthday.  The research of PLK has been supported by grants
from NSF and ARO.

\end{multicols}
\end{document}